\documentclass[journal=jacsat,manuscript=article]{achemso}

\usepackage{chemformula} 
\usepackage[T1]{fontenc} 

\usepackage{amsmath} 
\usepackage{physics} 
\usepackage{amssymb}
\usepackage{amsfonts}
\usepackage{slashed}
\usepackage{bbold}
\usepackage{xfrac}
\usepackage{bm}
\usepackage{siunitx}
\usepackage{subcaption}
\usepackage{pdfpages}

\author{Marit R. Fiechter}
\affiliation[ETH]
{Laboratory of Physical Chemistry, ETH Zürich, 8093 Zürich, Switzerland}
\author{Johan E. Runeson}
\affiliation[ox]
{Department of Chemistry, University of Oxford, Physical and Theoretical Chemistry Laboratory, South Parks Road, Oxford, OX1 3QZ, United Kingdom}
\author{Joseph E. Lawrence}
\affiliation[ETH]
{Laboratory of Physical Chemistry, ETH Zürich, 8093 Zürich, Switzerland}

\author{Jeremy O. Richardson}
\email{jeremy.richardson@phys.chem.ethz.ch}
\affiliation[ETH]
{Laboratory of Physical Chemistry, ETH Zürich, 8093 Zürich, Switzerland}

\title[An \textsf{achemso} demo]
  {How Quantum is the Resonance Behavior in Vibrational Polariton Chemistry?}

\begin{document}

\begin{tocentry}

\includegraphics{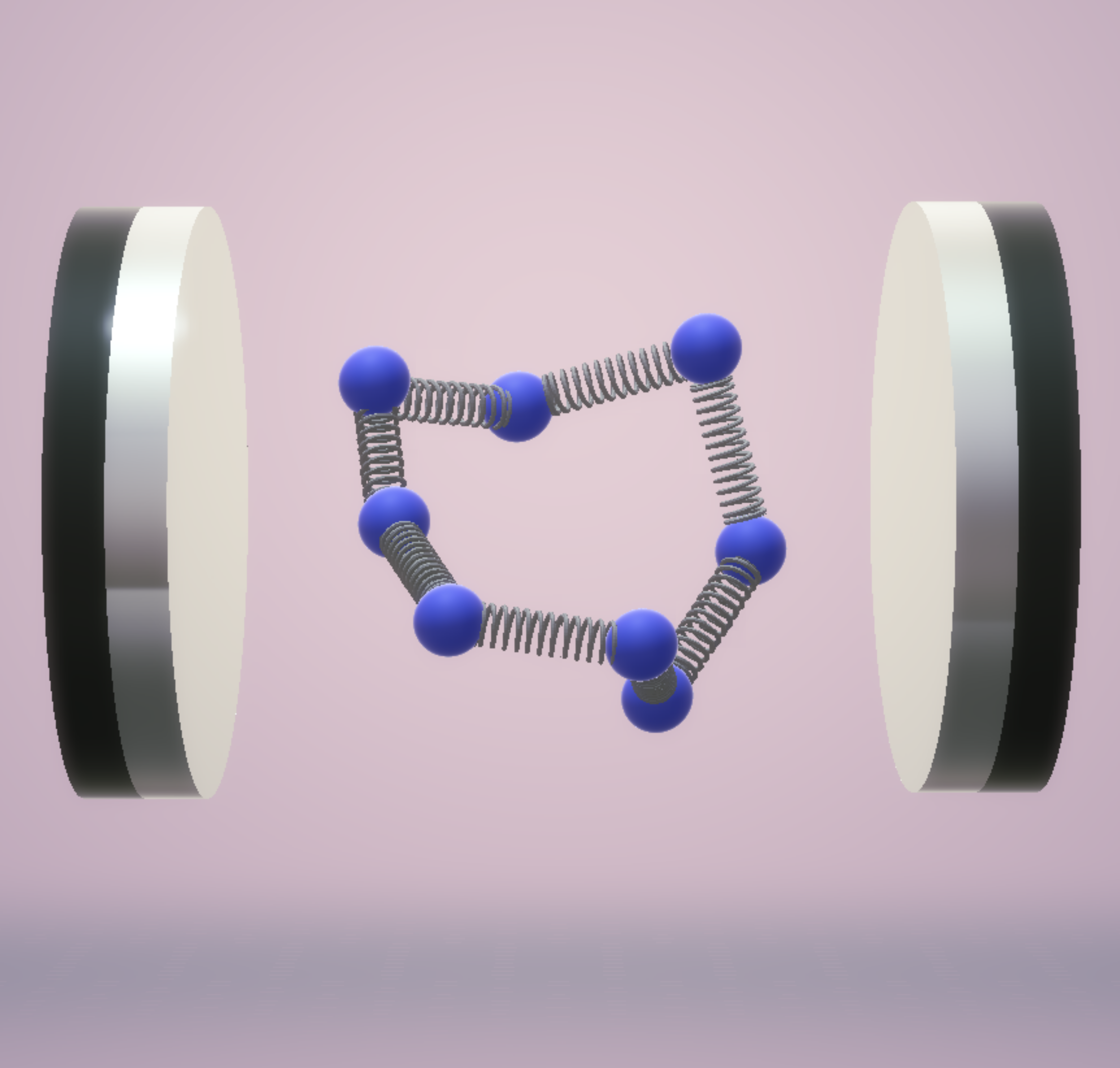}

\end{tocentry}

%

\begin{abstract}
   Recent experiments in polariton chemistry have demonstrated that reaction rates can be modified by vibrational strong coupling to an optical cavity mode. 
   Importantly, this modification only occurs when
   the frequency of the cavity mode is tuned to closely match a molecular vibrational frequency. This sharp resonance behavior has proved difficult to capture theoretically. Only recently, Lindoy et al.\ reported the first instance of a sharp resonant effect in the cavity-modified rate simulated in a model system using exact quantum dynamics. We investigate the same model system with a different method, ring-polymer molecular dynamics (RPMD), which captures quantum statistics but treats dynamics classically. We find that RPMD does not reproduce this sharp resonant feature at the well frequency, 
   and we discuss the implications of this finding for future studies in vibrational polariton chemistry. 
\end{abstract}

A recent series of experiments has revealed the surprising result that one can alter
chemical reaction rates just by placing the reaction mixture in an optical cavity \cite{hutchison2012modifying,thomas2016ground,thomas2019tilting,lather2019cavity,vergauwe2019modification,pang2020role,hirai2020modulation,sau2021modifying, ahn2023modification}, \emph{i.e.}~between a pair of carefully spaced mirrors which support standing waves of light at specific frequencies. 
In particular, when strongly coupling a cavity mode to molecular vibrations (called vibrational strong coupling, or VSC), \cite{nagarajan2021chemistry,dunkelberger2022vibration} the rate constant of ground-state reactions can be modified even without external driving, \emph{i.e.}\ without explicitly adding photons into the cavity. As the cavity mode can be treated as a harmonic oscillator coupled to the molecular system under study, it is relatively straightforward to incorporate into standard theoretical chemistry methods. However, in spite of the plethora of theoretical studies conducted on the topic, recently reviewed in \emph{e.g.}~Refs.~\citenum{ruggenthaler2022understanding,mandal2022theoretical,campos2022swinging}, the mechanism behind this cavity effect on the chemical reaction rate is not yet well understood. 

One of the features observed in experiments that has proven hard to reproduce theoretically is the resonance behavior: the rate-constant modification is only significant when the cavity length is tuned such that one of the cavity modes is in resonance with a vibrational mode in the reaction mixture (either of a reactant \cite{thomas2016ground,thomas2019tilting,hirai2020modulation,sau2021modifying,lather2019cavity,pang2020role,ahn2023modification} or a solvent molecule \cite{vergauwe2019modification,lather2019cavity,ahn2023modification}). The width of this resonant feature in the plot of rate vs.~cavity frequency is comparable to the linewidth of the molecular resonance in the infrared spectrum \cite{thomas2016ground,vergauwe2019modification,thomas2019tilting,hirai2020modulation}, typically on the order of tens of wavenumbers. Another feature of experiments complicating a theoretical analysis is the fact that, in experiments, a large number of molecules is coupled to a single cavity mode. This induces collective effects, so that spectral characteristics such as the Rabi splitting depend on the number of molecules coupled to the cavity \cite{casey2016vibrational,vergauwe2016quantum}. The question remains by what mechanism these collective effects influence the rate. However, for now we will constrain ourselves to the single-molecule regime, as the focus of this work is to compare an approximate theory to a fully quantum-mechanical benchmark, for which the extension to multiple molecules quickly becomes prohibitively expensive.


In this work, we focus on the resonance behavior and further investigate to what extent quantum effects could play a role for these sharp resonances in the rate around a reactant vibrational frequency. 
Although
analytical rate theories such as Grote--Hynes theory\cite{li2021cavity}, Eyring theory \cite{yang2021quantum} or Pollak--Grabert--H\"{a}nggi theory \cite{lindoy2022resonant} do predict a cavity effect, 
these effects tend to be spread over a broad range of cavity frequencies. Additionally, the largest effect observed in these theories does not necessarily occur when a cavity mode is on resonance with a vibrational mode, contrary to experimental observations.  
Only recently has a sharp resonance behavior, more in line with experiment, been reported theoretically by Lindoy \emph{et al.}\cite{lindoy2022quantum} In this study they used a fully quantum-dynamical approach (hierarchical equations of motion, HEOM) \cite{tanimura2020numerically} for a specific low-friction parameter regime of a model double-well system coupled to a cavity mode. These quantum simulations give peaks in the rate modification centered at the reactant frequency with a full-width half-maximum (FWHM) as small as \SI{80}{\per\centi\meter}, which is significantly narrower than the FWHM seen in earlier classical simulations \cite{sun2022suppression, wang2022cavity} (for example $\sim$\SI{350}{\per\centi\meter} for the isomerization of HONO on an \emph{ab initio} PES \cite{sun2022suppression}). Importantly, the quantum results are comparable to the resonance widths observed in experiment \cite{thomas2016ground,thomas2019tilting,vergauwe2019modification,ahn2023modification}, although it should be noted that Lindoy \emph{et al.}~considered only a single molecule in the cavity, making up for the lack of collective enhancement of the effect by choosing an extremely large light--matter coupling strength.

We assess the importance of quantum effects in forming these sharp peaks by comparing the HEOM results
with ring-polymer molecular dynamics (RPMD) rate theory \cite{craig2005chemical,craig2005refined,lawrence2020path}. RPMD is based on imaginary-time path integrals and is thus able to capture quantum statistics (including tunneling and zero-point energy effects), but treats dynamics classically, meaning it cannot reproduce effects due to quantum coherence.\cite{Althorpe2021Matsubara}
Hence, whether or not RPMD succeeds in reproducing the HEOM results tells us whether the quantum effects responsible for the sharp resonant feature are statistical or dynamical in nature. This will again have important implications for future studies in this field: as quantum-dynamics calculations on full-dimensional realistic systems are prohibitively expensive,
it is natural to resort to more affordable methods that capture quantum effects only approximately. \cite{chowdhury2021ring,sun2022suppression,schafer2022shining} 
The results in this study show that, if the single-molecule low-friction models studied in Ref.~\citenum{lindoy2022quantum} are indeed representative of experiment, one needs quantum dynamics to describe the physics correctly and reproduce sharp resonances. 

\emph{Model.} The model system we study here is equivalent to the model of Ref.~\citenum{lindoy2022quantum}. For completeness, in the following we give a full description of the form and parameters of this model.

\begin{figure}
     \centering
     \includegraphics[width=\textwidth]{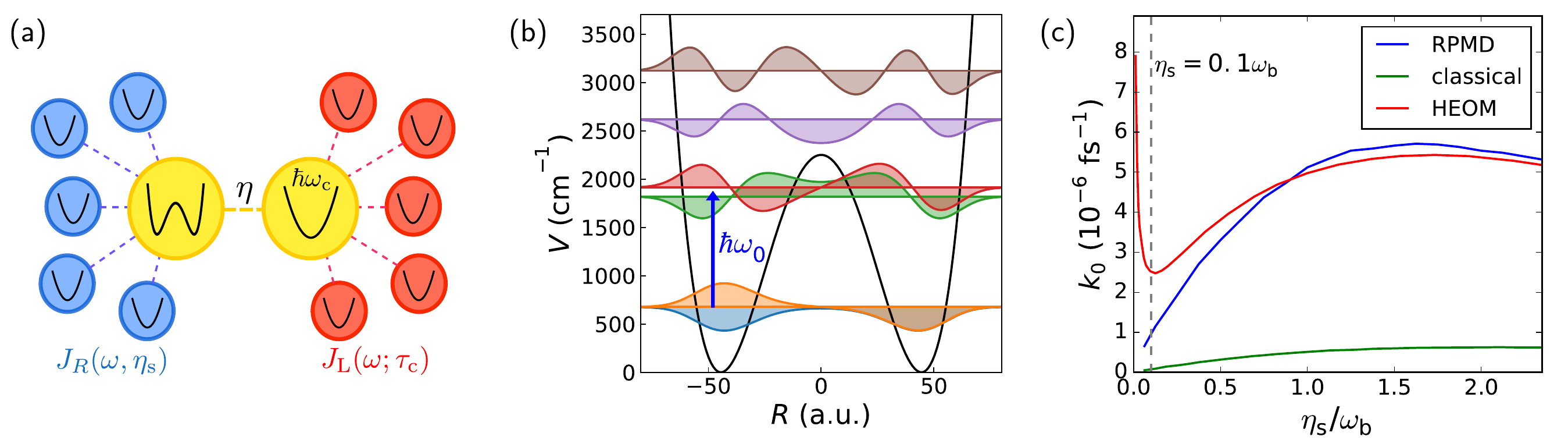}
        \caption{Outline of the model. (a) Diagram representing the degrees of freedom in the model and their coupling. The molecule-cavity coupling strength is given by $\eta$, the magnitude of the solvent friction determined by $\eta_\mathrm{s}$ (blue) and the strength of the coupling between cavity mode and external modes by $\tau_\mathrm{c}$ (red). (b) Double well \textbf{I} and its vibrational eigenstates, with the blue arrow indicating the well frequency $\omega_0$. (c) Rate constant $k_0$ as a function of solvent friction $\eta_\mathrm{s}$ outside the cavity (\emph{i.e.} $\eta=0$) for system \textbf{I}, with the gray dashed line indicating the friction at which the results of Figure~\ref{fig:shallow_cav} were obtained. }
        \label{fig:shallow}
\end{figure}

The model Hamiltonian can be expressed as 
\begin{equation}
    \hat{H} = \hat{H}_\text{mol} + \hat{H}_\text{solv} + \hat{H}_\text{cav} + \hat{H}_\text{cav-loss},
\end{equation}
where $\hat{H}_\text{mol}$ is the molecular Hamiltonian, $\hat{H}_\text{solv}$ describes the solvent and its coupling to the molecule, $\hat{H}_\text{cav}$ describes the cavity mode and its interaction with the molecule, and $\hat{H}_\text{cav-loss}$ represents cavity loss through interaction with electromagnetic modes outside the cavity. A diagrammatic representation of the degrees of freedom in this Hamiltonian and their coupling is shown in Figure~\ref{fig:shallow}a. All calculations are performed at a temperature of \SI{300}{\kelvin}.

The molecular Hamiltonian $\hat{H}_\text{mol}$ is chosen to be a one-dimensional symmetric double well, so that for the molecular coordinate $R$ (with unit mass) we have $\hat{H}_\text{mol}=\hat{P}_R^2/2 + V(\hat{R})$ with the potential given by
\begin{equation}
    V(R) = \frac{\omega_\mathrm{b}^4}{16 E_\mathrm{b}} R^4 - \tfrac{1}{2}\omega_\mathrm{b}^2 R^2 + E_\mathrm{b},
\end{equation}
where $\omega_\mathrm{b}$ is the imaginary part of the barrier frequency and $E_\mathrm{b}$ is the barrier height of the double well. 
Note that, as illustrated in Figure~\ref{fig:shallow}b, the vibrational states of such a double-well system are split by a very small energy, the tunneling splitting. We define the (anharmonic) well frequency as the difference between the first and second pair of these tunneling-split states, \emph{i.e.}~$\hbar\omega_0=\tfrac{1}{2}(E_2+E_3) - \tfrac{1}{2}(E_0+E_1)$, with $E_n$ denoting the energy of the $n$th eigenstate of $\hat{H}_\text{mol}$. In this work we will study two of these double-well systems: \textbf{I}, a relatively shallow double-well system (studied in the main text of Ref.~\citenum{lindoy2022quantum}), with $\omega_\mathrm{b} =\SI{1000}{\per\centi\meter}$ and $E_\mathrm{b} = \SI{2250}{\per\centi\meter}$, so that $\omega_0\approx \SI{1190}{\per\centi\meter}$; and \textbf{II}, a deeper double-well system (studied in the Supplementary Information of Ref.~\citenum{lindoy2022quantum}) with $\omega_\mathrm{b} =\SI{500}{\per\centi\meter}$ and $E_\mathrm{b} = \SI{2000}{\per\centi\meter}$, yielding $\omega_0\approx \SI{650}{\per\centi\meter}$.

The interaction of the molecule to the solvent is taken to be
\begin{equation}
\begin{aligned}
    \hat{H}_\text{solv} & = \sum_i \bigg[\frac{\hat{P}_i^2}{2} + \tfrac{1}{2}\Omega_i^2\bigg(\hat{Q}_i + \frac{C_i \hat{R}}{\Omega_i^2}\bigg)^2\bigg],
\end{aligned}
\end{equation}
which couples the molecular coordinate to a harmonic bath described by the canonical operators $\hat{P}_i$ and $\hat{Q}_i$. Here we take this bath to be characterized by a Debye spectral density $J_R(\omega) = \tfrac{\pi}{2}\sum_i \frac{C_i^2}{\Omega_i}\delta(\omega-\Omega_i) = \eta_\mathrm{s} \omega \Gamma^2 / (\omega^2 + \Gamma^2)$, where we set $\Gamma=\SI{200}{\per\centi\meter}$; $\eta_\mathrm{s}$ can be varied to change the solvent friction.

Within the dipole approximation, the coupling of the molecular coordinate to the cavity mode is given by the Pauli--Fierz Hamiltonian \cite{mandal2022theoretical,ruggenthaler2022understanding}, 
\begin{equation}
    \hat{H}_\text{cav} = \frac{1}{2}\hat{p}_\mathrm{c}^2 + \frac{1}{2}\omega_\mathrm{c}^2 \left(\hat{q}_\mathrm{c}+\sqrt{\frac{2}{\hbar \omega_\mathrm{c}}}\eta \hat{\mu}\right)^2.
\end{equation}
Here, $\hat{p}_\mathrm{c}$ and $\hat{q}_\mathrm{c}$ are the canonical displacement field operators, $\omega_\mathrm{c}$ is the cavity frequency, and $\eta=\sqrt{\frac{\hbar}{2\omega_\mathrm{c} \epsilon_0 V}}$ is a measure of the coupling strength, with $\epsilon_0$ the permittivity of vacuum and $V$ the quantization volume. $\hat{\mu}$ is dipole moment operator projected onto the electronic ground state and along the cavity polarization, $\hat{\mu}=\hat{\mathbf{e}}\cdot\hat{\boldsymbol{\mu}}$. We follow Ref.~\citenum{lindoy2022quantum} and choose the dipole moment to be linear in the molecular coordinate, $\hat{\mu}=\hat{R}$.

If the cavity mirrors are not perfectly reflective, the electromagnetic mode inside the cavity can interact with the continuum of modes outside the cavity, enabling for example the escape of a photon from the cavity. This can effectively be described by
\begin{equation}
    \hat{H}_\text{cav-loss} = \sum_j \bigg[ \frac{\hat{p}_j^2}{2} + \frac{1}{2}{\omega}_j^2\bigg(\hat{q}_j + \frac{c_j \hat{q}_\mathrm{c}}{{\omega}_j^2}\bigg)^2\bigg],
\end{equation}
where $\hat{{p}}_j$ and $\hat{{q}}_j$ are the canonical operators associated to field modes outside the cavity.
Making the assumption that this bath of external light modes is Markovian, one can relate the parameters in this Hamiltonian, \emph{i.e.}~the set of frequencies ${\omega}_j$ and couplings $c_j$ described by a spectral density $J_\mathrm{L}(\omega)=\frac{\pi}{2}\sum_j\frac{c_j^2}{{\omega}_j}\delta(\omega-{\omega}_j )$, to the cavity lifetime $\tau_\mathrm{c}$. One possible such relation is given in Ref.~\citenum{lindoy2022quantum}:
\begin{equation} \label{eq:lifetimeJ}
    \tau_\mathrm{c}=\omega_\mathrm{c}\frac{1-\mathrm{e}^{-\beta\hbar\omega_\mathrm{c}}}{2J_\mathrm{L}(\omega_\mathrm{c})},
\end{equation}
which is the definition we will use throughout this work for consistency.\footnote{We note that alternatively, one could take $\tau_\mathrm{c}=\omega_\mathrm{c}/J_\mathrm{L}(\omega_\mathrm{c})$; this relation can be obtained by calculating the expectation value of the photon number operator for an empty cavity (no matter) with the bath at zero temperature, giving an exponential decay that allows for extraction of a lifetime: $\langle \hat{n}(t) \rangle=\langle \hat{n}(0) \rangle \mathrm{e}^{-J(\omega_\mathrm{c}) t/\omega_\mathrm{c}}  $ (Ref.~\citenum{carmichael1999statistical}, Eq.~1.80). }
In Ref.~\citenum{lindoy2022quantum}, $J_\mathrm{L}(\omega)$ was chosen to be a Debye spectral density $J_\mathrm{L}(\omega) = \eta_\mathrm{L} \omega \Gamma_\mathrm{L}^2 / (\omega^2 + \Gamma_\mathrm{L}^2)$ with $\Gamma_\mathrm{L}=\SI{1000}{\per\centi\meter}$. 
Debye baths are the natural choice for HEOM calculations, whereas for RPMD, there is a significant advantage in using an Ohmic spectral density instead.
Therefore, to make the cost of our RPMD calculations tractable for system \textbf{I}, we replaced this Debye spectral density by an Ohmic spectral density $J_\mathrm{L}(\omega)=\gamma_\mathrm{L}\omega$ with $\gamma_\mathrm{L}$ defined by $\tau_\mathrm{c}$ and Eq.~\ref{eq:lifetimeJ}. 
As the spectral density near $\omega_\mathrm{c}$ is unaltered, this is not expected to significantly change the dynamics.
Details on our treatment of the spectral densities and results supporting the validity of this exchange are given in the Supporting Information.

\newpage
\emph{Theory. }
We perform the calculations in this study with RPMD \cite{craig2004quantum,craig2005chemical,craig2005refined,habershon2013ring}. Here we  summarize the general idea behind this method; for details of our implementation, we refer the reader to the Supporting Information.

In short, RPMD is based on  discretized closed paths in imaginary time called ring polymers. These ring polymers emerge for example in path-integral molecular dynamics, \cite{parrinello1984study,marx1996ab} where, by sampling over ring-polymer configurations, one can compute static equilibrium properties of a quantum system exactly. It is however not feasible to rigorously extend this to real-time dynamics because of the infamous sign problem. RPMD circumvents this by instead propagating the ring polymer classically in time to obtain an approximation to real-time quantum correlation functions such as the flux--side correlation function, from which one can extract the rate constant. 
RPMD rate theory can be shown to capture static quantum effects such as zero-point energy effects and tunneling \cite{craig2005chemical,craig2005refined,richardson2009ring,Hele2013QTST,lawrence2020path}; on the other hand it is not able to capture truly dynamical quantum effects, such as interference. \cite{Althorpe2021Matsubara}






\emph{System I: shallow wells.}  We start by investigating the behavior of the chemical reaction rate without the cavity, $k_0$, with increasing solvent friction, $\eta_\mathrm{s}$, in Figure~\ref{fig:shallow}c. For frictions larger than $\eta_\mathrm{s}\approx0.2\omega_\mathrm{b}$, this figure shows typical Kramers turnover behavior \cite{peters2017reaction16,craig2005chemical}, i.e.~the rate increases with increasing friction until it reaches a maximum (``turnover''), and then decreases with increasing friction after that. For very low friction however, the exact rate in this system shoots up again due to coherent nuclear tunneling between states localized in the left and in the right well; this is a dynamic quantum effect not captured by RPMD. For higher frictions, the RPMD rate more accurately approximates the quantum rate, being within a few percent after the Kramer's turnover. The classical rate on the other hand underestimates the quantum benchmark rate by an order of magnitude.



\begin{figure}[h]
    \centering
    \includegraphics[width=\textwidth]{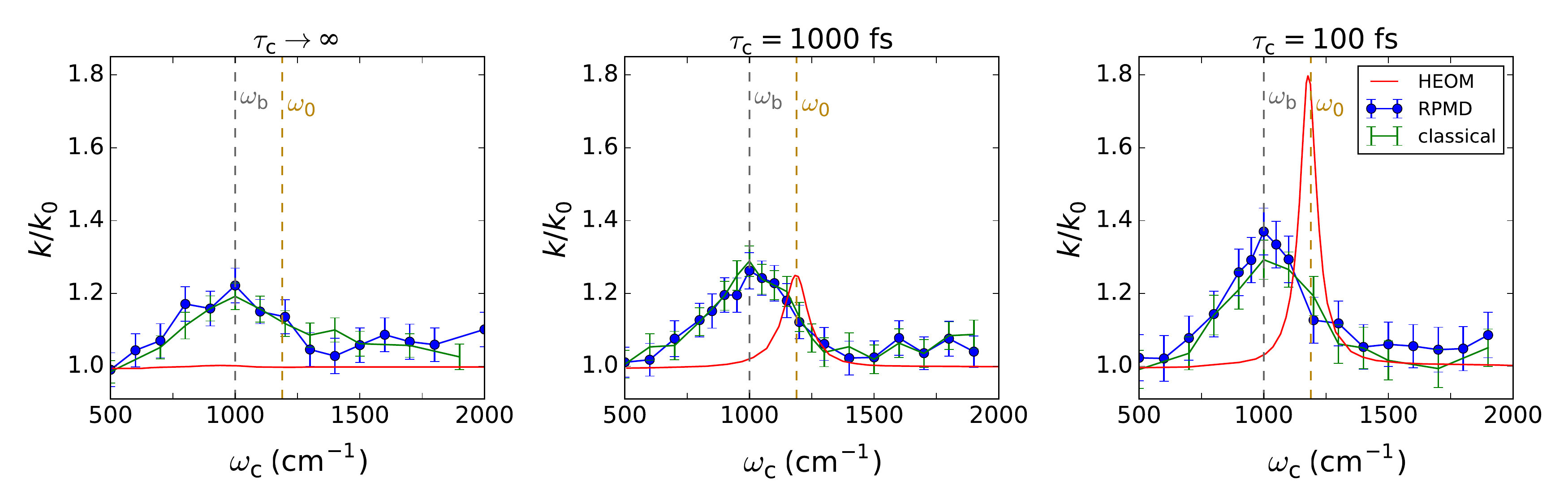}
    \caption{Cavity modification of the rate of system \textbf{I} as a function of cavity frequency $\omega_\mathrm{c}$ for a set of cavity lifetimes $\tau_\mathrm{c}$, with $\eta=0.00125$ a.u.~and $\eta_\mathrm{s}=0.1\omega_\mathrm{b}$. Neither classical dynamics nor RPMD is able to reproduce the sharp peak in rate enhancement at the well frequency $\omega_0$; instead they give a broad peak centered at the barrier frequency $\omega_\mathrm{b}$. The error bars indicate a 68\% confidence interval. }
    \label{fig:shallow_cav}
\end{figure}

In Figure~\ref{fig:shallow_cav} we investigate the change in rate, $k/k_0$, when coupling the molecular coordinate to the cavity mode for a low value of the solvent friction, $\eta_\mathrm{s}=0.1\omega_\mathrm{b}$. In particular, in Ref.~\citenum{lindoy2022quantum} it was found that the effect of the cavity is negligible when the cavity is lossless ($\tau_\mathrm{c}\rightarrow\infty$): in that case energy transfer from the molecular mode to the cavity mode may be possible, but from there the energy cannot dissipate. The cavity mode will therefore not efficiently dampen the motion in the reaction coordinate. 
When increasing the coupling between the cavity mode and the environment (\emph{i.e.} reducing $\tau_\mathrm{c}$), a dramatic increase in the exact rate is observed. Moreover, the resulting rate profile as a function of cavity frequency features a single sharp peak (with a full-width half-maximum (FWHM) of about $\SI{80}{\per\centi\meter}$ for $\tau_\mathrm{c}=\SI{100}{\femto\second}$ cavity lifetime), and it is centered around the well frequency. At least in this respect it is in agreement with experimental observations (although the experiments involve many molecules in a cavity, and it is unclear whether the current model is physically realistic for a reaction in solution). 

Both classical and RPMD simulations display a markedly different behavior from the HEOM results: the peak in the cavity-induced rate enhancement is centered around the barrier frequency $\omega_\mathrm{b}$ rather than the well frequency, and it is also a much broader feature, its FWHM spanning hundreds of wavenumbers. Interestingly, the effect of cavity loss ($\tau_\mathrm{c}$) on the rate is much smaller here: the rate modification is not entirely suppressed for lossless cavities, as it was in the quantum case, and the results do not change dramatically when introducing a finite cavity lifetime (from a rate enhancement by a factor of 1.2 for a lossless cavity to a factor of 1.4 for a cavity with lifetime $\tau_\mathrm{c}=\SI{100}{\femto\second}$).

Even though the absolute rate without cavity, $k_0$, predicted by classical and RPMD simulations differs by a factor of about $8$, they predict similar rate modifications, $k/k_0$, which agree with each other within the error bars. This implies that quantum statistics does not play an important role in the effect the cavity has on the relative rate $k/k_0$ in this case. On the other hand, the cavity modification of the RPMD rate constant does not even qualitatively agree with the full quantum rate constant, implying the quantum dynamics is crucial in this model system for capturing the right behavior.

\begin{figure}
     \centering
     \includegraphics[width=\textwidth]{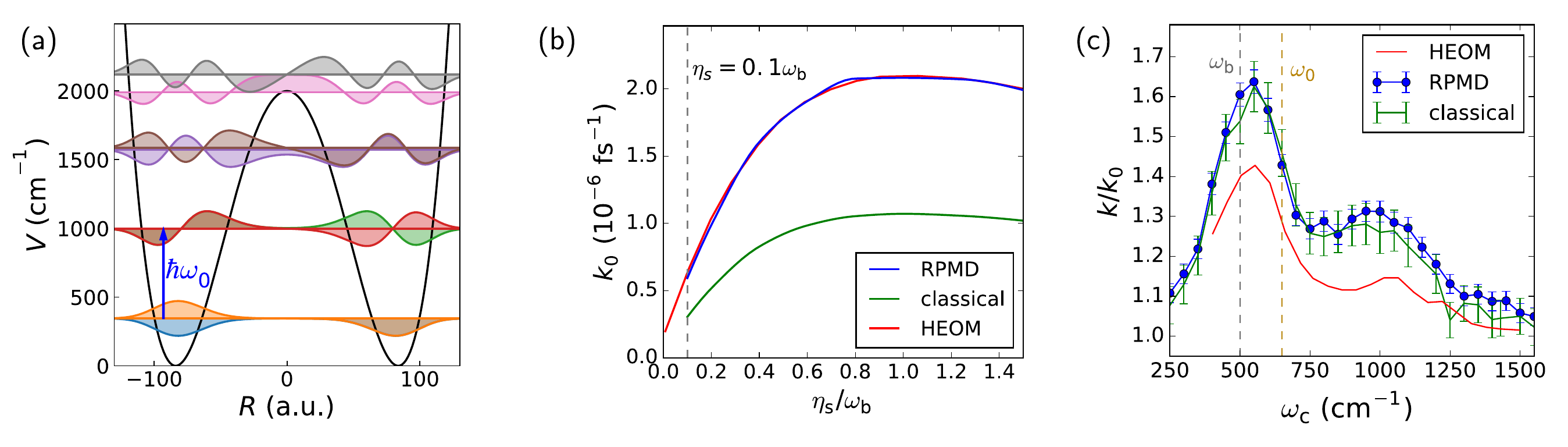}
        \caption{System \textbf{II} and cavity modification of its rate. (a) Double well \textbf{II} and its eigenstates, with the blue arrow indicating the well frequency $\omega_0$. (b) Rate constant $k_0$ as a function of solvent friction $\eta_\mathrm{s}$ outside the cavity ($\eta=0$). (c) Rate-constant modification $k/k_0$ as a function of cavity frequency for $\eta_\mathrm{s}=0.1\omega_\mathrm{b}$, $\eta=0.005$ a.u., and $\tau_\mathrm{c}=1000$ fs.}
        \label{fig:deep}
\end{figure}

\emph{System II: deep wells.} We now move on to a double-well system with a higher barrier, so that there are three (instead of two) tunneling-split eigenstates below the barrier (see Figure \ref{fig:deep}a, \emph{c.f.}\ Figure \ref{fig:shallow}b). The cavityless rate constant as a function of solvent friction $\eta_\mathrm{s}$, shown in Figure~\ref{fig:deep}b, reveals that this system loosely speaking behaves less quantum mechanically; there is for example no coherent-tunneling regime at very low frictions. The classical rate in this case only underestimates the full quantum rate constant by a factor of 2, whereas RPMD yields a spot-on prediction of the rate.

Coupling the cavity mode to this deeper double well yields quite a different cavity-frequency-dependence of the rate, as displayed in Figure~\ref{fig:deep}c. Firstly, the rate profile is not composed of a single sharp peak; rather, it is made up of a much broader peak (FWHM $>\SI{300}{\per\centi\meter}$) and a somewhat lower broad shoulder. Moreover, the main peak is no longer centered at the molecular vibrational frequency; at $\omega_\mathrm{c}\approx\SI{560}{\per\centi\meter}$ it is somewhere in between the barrier frequency ($\omega_\mathrm{b}=\SI{500}{\per\centi\meter}$) and the well frequency ($\omega_0\approx\SI{650}{\per\centi\meter}$). 

Interestingly, in this case the rate-enhancement profile is captured qualitatively by the classical simulations, and adding in quantum statistics via RPMD has little additional effect. This suggests that the origin of the cavity-induced rate modification in this system can be understood classically. Note, however, that by going from system \textbf{I} to system \textbf{II}, we have also lost the sharp peak centered at $\omega_0$, and therefore this system does not exhibit one of the key features of the experimental results. 

It should be noted, however, that although RPMD agrees very well with HEOM in the cavityless case, it overestimates the magnitude of the rate modification by the cavity. This is in line with the saturation of the HEOM rate modification for larger light--matter coupling strengths, $\eta$, observed in the Supplementary Information of Ref.~\citenum{lindoy2022quantum}; this saturation effect is much less pronounced in the classical case. RPMD does not improve upon classical dynamics in capturing this, suggesting the saturation effect is due to a mechanism based on coherent exchange of energy between the cavity mode and the double-well system.  

\emph{Discussion. }
In summary, we have investigated two variations of the model system studied in Ref.~\citenum{lindoy2022quantum}. In system \textbf{I}, where the molecular coordinate is a relatively shallow double well, full quantum dynamics predicts that the rate modification as a function of cavity frequency is peaked sharply around the well frequency. We show that this feature cannot be reproduced with classical dynamics, and that even adding quantum statistics with RPMD does not improve upon this. This indicates that the resonance in this system requires quantum dynamical effects.  In system \textbf{II} we consider a deeper double-well system, where intuitively quantum effects should play a less prominent role. Indeed, even classical dynamics qualitatively captures the correct frequency-dependent rate modification $k/k_0$ in this case (even though one needs statistical quantum effects in RPMD to correctly predict the absolute rate constant). This system however lacks the provocatively sharp resonance feature that made system \textbf{I} so intriguing. 
Also note that, surprisingly, introducing coupling to the cavity mode causes RPMD to overestimate the rate by $\sim$14\%, even though it was spot-on in the cavityless case.


The implications of these findings are as follows. We have seen that the only sharp peak in the rate modification predicted in theory so far can exclusively be simulated by quantum dynamics. However, this has only been observed for a specific model, and it is not at all clear whether this model is representative of the reactions done in experiment. Firstly, in this model, the molecular coordinate is only very weakly coupled to its bath, so that the reaction takes place in the low-friction regime. Chemical reactions typical of this regime are unimolecular dissociation reactions in low-pressure gases \cite{peters2017reaction16}; in this light, it seems unlikely that the model studied here is a good presentation of the (solution-phase) reactions performed in experiment.
Secondly, our findings seem to suggest that 
only systems for which the rate exhibits quantum-dynamical behavior (even before coupling to the cavity) will exhibit sharp resonances.
This is rather atypical for solution-phase reactions, where coherences are expected to wash out rapidly \cite{habershon2013ring}. In fact, this is the reason that RPMD normally performs so well for predicting reaction rates in solution; the very fact that RPMD fails to correctly predict the rate of system \textbf{I} at the friction of interest without cavity coupling in itself already suggests that this system may not be representative of reactions in solution.

Nevertheless, it cannot be ruled out that somehow, in the reactions studied in experiment, a critical part of the motion leading to products is only weakly damped, effectively introducing a low friction locally and leading to a situation in which by some means coherences can survive longer than expected. 
Even if this is true, and quantum dynamics is vital for capturing the resonance behavior correctly, not all is lost; one may for example still be able to cut computational costs by only treating important molecular and cavity degrees of freedom quantum mechanically in a mixed quantum--classical scheme \cite{tully1990FSSH,meyera1979classical,stock1997semiclassical,runeson2019spin,runeson2020generalized,ultrafast, mannouch2022mapping, runeson2022fmo}. 

Another possibility is that the sharp peak in rate modification in system \textbf{I} is more of an exotic feature, the double well being very shallow, and the resonance prominent at a friction where the rate is enhanced by coherent nuclear tunneling. The sharp peaks observed in experiment, however, may be of a different kind. That they have not been observed in simulations could be because our models or methods in the high friction regime just still lack essential features.

Finally, it is also possible that by some unknown mechanism one can only get these resonances to be sharp in the high-friction regime by accounting for collective effects. This has previously been suggested in for example Refs.~\citenum{schafer2022shining,sidler2022perspective,ahn2023modification}, and is an interesting avenue for further exploration. 



\paragraph{Acknowledgements} The authors would like to thank Lachlan Lindoy for helpful discussions. J.E.R. would like to thank David Manolopoulos for advice on evolving the explicit bath.
M.R.F. was supported by an ETH Zurich Research Grant,  J.E.L. was supported by an ETH Zurich Postdoctoral Fellowship, and J.E.R. by a mobility fellowship from the Swiss National Science Foundation.

\paragraph{Supporting Information Available}
The following files are available free of charge:
\begin{itemize}
    \item SupportingInformation.pdf: details on the computational method used and results supporting the validity of the replacement of the Debye spectral density by an Ohmic spectral density in the light-mode bath.
\end{itemize}

\bibliography{ref.bib}
\includepdf[pages={1-5}]{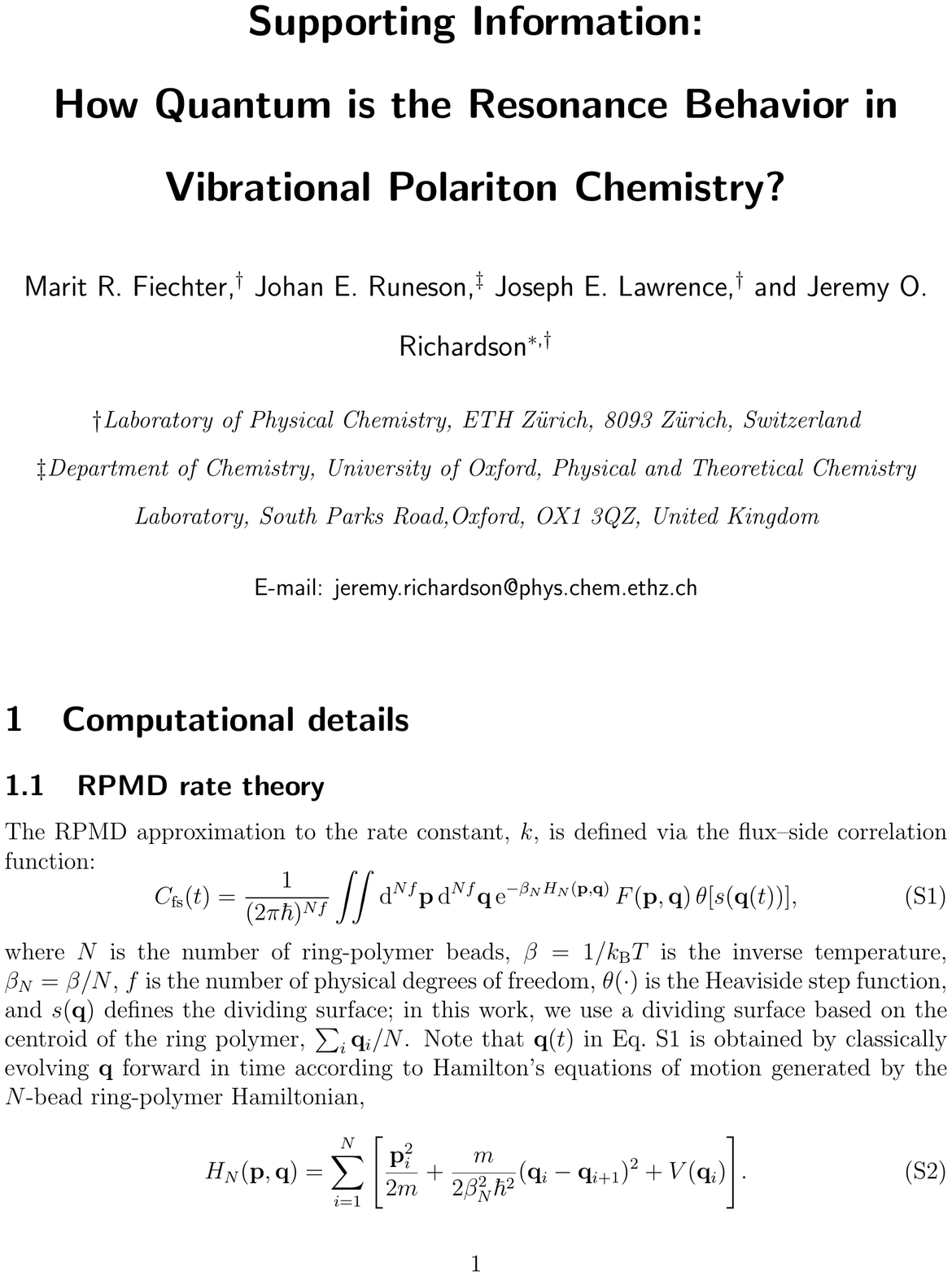}

\end{document}